\documentstyle[12pt,epsfig]{article}

\newcommand{\spa}{\vspace{.25cm}}
\newcommand{\beq}{\begin{equation}}
\newcommand{\eeq}{\end{equation}}
\newcommand{\beqs}{\begin{eqnarray}}
\newcommand{\eeqs}{\end{eqnarray}}
\newcommand{\bary}{\begin{array}}
\newcommand{\eary}{\end{array}}
\newcommand{\nid}{\noindent}
\newcommand{\sni}{\spa \nid}

\newcommand{\x}{\ell}
\newcommand{\SN}{solar neutrino }
\newcommand{\FCNC}{FCNCs }
\newcommand{\step}{\hspace{1.4cm}}

\newcommand{\figpos}{p}        
\newcommand{\spaAR}{7.5}       
\newcommand{\heightAR}{16.2}   
\newcommand{\widthAR}{16.2}    
\newcommand{\R}{\bigr>}
\newcommand{\B}{\bigr<}
\newcommand{\K}{{\cal{K}}}

\newcommand{\Op}{{\cal{O}}}
\newcommand{\bU}{\mbox{\boldmath $U$}}
\newcommand{\bM}{\mbox{\boldmath $M$}}
\newcommand{\bPsi}{\mbox{\boldmath $\Psi$}}

\newcommand{\APJ}{Astrophys. J. }
\newcommand{\PRD}{Phys. Rev. D }
\newcommand{\PRL}{Phys. Rev. Lett. }
\newcommand{\PLB}{Phys. Lett. B }
\newcommand{\NPB}{Nucl. Phys. B }

\def\gsim{\ \rlap{\raise 3pt \hbox{$>$}}{\lower 3pt \hbox{$\sim$}}\ }
\def\lsim{\ \rlap{\raise 3pt \hbox{$<$}}{\lower 3pt \hbox{$\sim$}}\ }

\def\putMSW#1#2#3#4#5 
{
 \begin{figure}[\figpos]
 \LARGE
 $\Delta$ [eV$^2$]
 \begin{center}
 \mbox{\epsfig{figure=#1.eps,angle=0,width=#3cm,height=#4cm}}
 \end{center}
 \hspace{#5cm}
 $\sin^2 2\theta$
 \spa
 \caption{#2}
 \label{#1}
 \end{figure}
}

\begin{document}

\title{$Z$-induced \FCNC and their effects on\\ Neutrino Oscillations}

\author{Sven Bergmann$^a$ and Alex Kagan$^b$\\
        $^a${\small \it Department of Particle Physics,
        Weizmann Institute of Science,
        Rehovot 76100, Israel}\\
        $^b${\small \it Department of Physics, 
        University of Cincinnati, 
        Cincinnati OH 45221}}

\date{(WIS-98/4/MAR-DPP)}

\maketitle

\begin{abstract}%
\sni%
Adding singlet neutrinos to the standard model spectrum in general 
gives rise to $Z$-induced flavor-changing neutral currents.
We study the impact of these currents on matter-induced neutrino
oscillations in the sun and in supernovae.  While the effects for
solar neutrinos are negligible, dramatic effects are possible 
for supernova neutrinos.
\end{abstract}%


\section{Introduction}
Neutrino oscillations~\cite{KuoPanta} can provide an appealing
solution to the solar neutrino problem, which seems to be increasingly
difficult to explain otherwise.  All four \SN experiments
\cite{HS,GALLEX,SAGE,KK} observe a deficit of $\nu_e$'s compared to
the Standard Solar Model (SSM)~\cite{SSM} expectation.  In particular,
the apparently strongest depletion of intermediate energy neutrinos
makes it very difficult if not impossible to account for these
observations in {\it any} reasonable solar model \cite{Hata}.
Moreover the recent helioseismological data confirms the SSM
predictions for the solar density profile \cite{helioseis}
strengthening the belief that the \SN problem is due to non-standard
neutrino properties.

\spa%
The features that are required to allow for neutrino oscillations are
a non-vanishing mass-squared difference $\Delta_{ij} \equiv m_i^2-m_j^2$
(implying that at least one neutrino state is massive) and mixing
(i.e. the neutrino interaction eigenstates do not coincide with the
mass eigenstates).  The \SN problem can then be solved by vacuum or
matter-enhanced neutrino oscillations~\cite{MSW} with

\beq
\Delta_{sol}^{vac} \simeq 10^{-10} ~\mbox{eV}^2 ~~~~~\mbox{or}~~~~~
\Delta_{sol}^{mat} \simeq 10^{-5} ~\mbox{eV}^2.
\eeq

\sni%
The upper bounds~\cite{PDG} on the light neutrino masses,

\beq
m_{\nu_e}    \leq 15\  \mbox{eV},\ \ \
m_{\nu_\mu}  \leq 0.17\ \mbox{MeV},\ \ \
m_{\nu_\tau} \leq 24\ \mbox{MeV},
\eeq

\sni%
are not in conflict with such a solution, although one has to explain
why the postulated neutrino masses are so much smaller than the
charged fermion masses.

\spa%
The most popular solution is the Majorana {\it see-saw}
mechanism~\cite{seesaw}. It requires additional right-handed $SU(2)_L$
singlet fermions with large Majorana masses which mix with the Standard 
Model (SM) neutrinos via $SU(2)_L$ breaking Dirac masses.  In its
most common version this leads to a $6 \times 6$ mass matrix of the
form

\beq
\bM = \pmatrix{0         & \bM_D \cr
               \bM^T_D   & \bM_R}.
\label{see-saw-matrix}
\eeq

\sni%
The entries of the Dirac mass matrix $\bM_D$ are typically comparable
to the masses of the charged leptons while the entries of the Majorana
mass matrix $\bM_R$ do not break weak $SU(2)_L$ and are therefore
related to some very high or intermediate New Physics (NP) scale
$\Lambda$. This leads to three ultra-light neutrinos with masses of
order $m_D^2/m_R$ ($m_X \equiv (\mbox{det} \bM_X)^{1/3}$), and singlet
admixtures in the light mass eigenstates of order $m_D/m_R $. There
are variants of the above see-saw mechanism in which expanded neutrino
singlet sectors lead to much larger mixing with the known neutrinos.
We return to a discussion of these scenarios later in the paper.

\spa%
The point we wish to make is the following: In order to explain the
\SN problem (and other experimental anomalies like the atmospheric
neutrino problem~\cite{ANproblem} and the LSND
results~\cite{LSNDdar,LSNDdif}) one requires small, but non-zero
neutrino masses.  The most attractive scenarios for obtaining such
tiny masses involve see-saw suppression. The important ingredient is
the presence of heavy $SU(2)_L$ singlet neutrinos, which is rather
common in NP models.  However, mixing of the known neutrinos with
singlets may also influence the neutrino oscillations that were
invoked to solve the experimental anomalies in the first place, since
they modify the neutral current interactions in a way that is flavor
dependent. While the existence of this effect is generic in models
that allow for neutrino masses, it has not been extensively
investigated in the literature~\cite{Z-FCNC}. It is the purpose of
this work to give a quantitative analysis of its importance for solar
and supernova neutrinos. We find that phenomenological constraints on
singlet mixing make its significance for solar neutrinos marginal at
best, but allow for dramatic effects in the case of supernova
neutrinos.

\spa%
The paper is structured as follows: In Section~\ref{formalism} we
introduce the formalism of $Z$-induced Flavor Changing Neutral
Currents (FCNCs) and work out the changes to matter-enhanced neutrino
oscillations in the presence of non-sequential neutrinos. 
Phenomenological constraints on neutrino-singlet mixing are discussed 
in Section~\ref{constraints}. The implications for the
Mikheyev-Smirnov-Wolfenstein (MSW) solution of the \SN problem are
studied in Section~\ref{sun} and for oscillations of supernova
neutrinos in Section~\ref{supernova}.  We conclude in
Section~\ref{conclusions}.


\section{Formalism}
\label{formalism}

\subsection{$Z$-mediated FCNCs}
Consider a model where the lepton sector of the SM is extended in a
non-sequential way (we first consider additional $SU(2)_L$ singlets,
then briefly discuss the case of additional triplets in
Section~\ref{triplets}). We group the known ($K$) and the new ($N$)
interaction eigenstates in the vector $\bPsi^I = (\bPsi_K,
\bPsi_N)^T$, which is related to the corresponding vector of light
($L$) and heavy ($H$) mass eigenstates $\bPsi^M = (\bPsi_L,
\bPsi_H)^T$ by a unitary transformation

\beq
\pmatrix{\bPsi_K \cr \bPsi_N} = \bU \pmatrix{\bPsi_L \cr \bPsi_H}
\step \mbox{where} \step
\bU=\pmatrix{\bU_{K L} & \bU_{K H} \cr
             \bU_{N L} & \bU_{N H} }.
\eeq

\sni%
The submatrices $\bU_{K L}$ and $\bU_{N L}$ describe the overlap of
the light eigenstates with the known interaction states and the new
interaction states, respectively.

\spa%
It has been known for a long time \cite{Schechter} that the presence
of new interaction eigenstates in general leads to FCNC neutrino
interactions.  To make this statement more precise consider the
Neutral Current (NC) operator

\beq
\Op_{NC} \equiv |\Psi_\K \R \B \Psi_\K|,
\label{NCop}
\eeq

\sni%
where the sum over $\K = e, \mu, \tau$ is implicit.  Effective
four-Fermi couplings for $Z$-mediated neutrino scattering are
proportional to $\Op_{NC}$.  Trivially, the matrix elements of this
operator in the basis of known interaction eigenstates $\{\Psi_\K\}$
are

\beq
\B \Psi_\K |\Op_{NC}| \Psi_{\K'} \R = \delta_{\K \K'},
\eeq

\sni%
while in the basis of {\it light} mass eigenstates they are

\beq
\B \Psi_l |\Op_{NC}| \Psi_{l'} \R =
\B \Psi_l|\Psi_\K \R \B \Psi_\K|\Psi_{l'} \R =
U_{\K l} U^*_{\K l'}.
\label{O_l}
\eeq

\sni%
The latter are in general not equal to $\delta_{l l'}$, since the
submatrix $\bU_{K L}$ is not unitary. Thus there are
``flavor''-changing NCs between the light mass eigenstates.

\spa%
In most cases of interest the heavy mass eigenstates are not
kinematically accessible. Hence we need to project the interaction
eigenstates onto the subspace of light ``propagating'' mass eigenstates, if
we want to calculate the effective operators at low energies.  We
denote the projected states by

\beq
|\Psi_\K^P \R \equiv |\Psi_{l} \R \B \Psi_l| \Psi_\K \R =
U_{\K l} |\Psi_l \R.
\eeq

\sni%
Note that these states are not orthonormal to each other, since

\beq
\B \Psi_\K^P|\Psi_{\K'}^P \R = 
\B \Psi_\K|\Psi_{l} \R \B \Psi_l|\Psi_{\K'} \R = 
U^*_{\K l}  U_{\K' l}.
\eeq

\sni%
In the basis of these ``light'' interaction eigenstates the operator
$\Op_{NC}$ is represented by the matrix

\beq
\B \Psi_\K^P |\Op_{NC}| \Psi_{\K'}^P \R =
\B \Psi_\K|\Psi_{l} \R \B \Psi_l|\Psi_{\K''} \R
\B \Psi_{\K''}|\Psi_{l'} \R \B \Psi_{l'}|\Psi_{\K'} \R =
U^*_{\K l} U_{\K'' l} U^*_{\K'' l'} U_{\K' l'},
\label{O_P}
\eeq

\sni%
which is not in general diagonal since the basis $\{\Psi_l\}$ does not
span the space on which $\Op_{NC}$ is defined.

\spa%
As a simple example consider the case of two known interaction
eigenstates $\nu_e$ and $\nu_\x$ ($\x$ =
$\mu$ or $\tau$) and one new $SU(2)_L$ singlet $\nu_S$. The
vector $\bPsi^I = (\nu_e, \nu_\x, \nu_S)^T$ is connected to the vector
of mass eigenstates $\bPsi^M = (\nu_1, \nu_2, \nu_h)^T$, where $\nu_1$
and $\nu_2$ are light while $\nu_h$ is heavy, by a unitary
transformation

\beq
\pmatrix{\nu_e \cr  \nu_\ell \cr  \nu_S} =
\pmatrix{U_{e1}  & U_{e2}  & U_{e h}  \cr
         U_{\x 1} & U_{\x 2} & U_{\x h} \cr
         U_{S1}  & U_{S2}  & U_{S h} }
\pmatrix{\nu_1 \cr  \nu_2 \cr  \nu_h}.
\eeq

\sni%
The ``propagating'' neutrinos that are produced in low-energy
charged-current interactions together with the charged leptons $e$ and
$\x$ are

\beq
\pmatrix{\nu_e^P \cr  \nu_\x^P} =
\pmatrix{U_{e1}  & U_{e2}  \cr
         U_{\x 1} & U_{\x 2} }
\pmatrix{\nu_1 \cr  \nu_2},
\eeq

\sni%
i.e. we have projected $\nu_e$ and $\nu_\x$ onto the $\nu_1-\nu_2$
plane.  Note that since $\bU_{K L}$ is not unitary, $\nu_e^P$ and
$\nu_\x^P$ are {\it not} orthogonal to each other,

\beq
\B \nu_e^P | \nu_\x^P \R = U_{e1}^* U_{\x 1} + U_{e2}^* U_{\x 2}
                         = - U_{eh}^* U_{\x h},
\label{ortho}
\eeq

\sni%
and are not properly normalized. Thus we cannot use these states as a
proper basis for the description of neutrino oscillations. Instead, we
choose an orthonormal basis $\{|\nu_e^O\R, |\nu_\x^O\R\}$ where

\beq
|\nu_e^O\R = \frac{U_{e1} |\nu_1\R + U_{e2} |\nu_2\R}
                  {\sqrt{|U_{e1}|^2+|U_{e2}|^2} }
\eeq

\sni%
is aligned with $|\nu_e^P\R$ and

\beq
|\nu_\x^O\R = \frac{-U_{e2}^* |\nu_1\R + U_{e1}^* |\nu_2\R}
                   {\sqrt{|U_{e1}|^2+|U_{e2}|^2} }
\eeq

\sni%
is orthogonal to $|\nu_e^O\R$. In this basis the off-diagonal matrix elements
of $\Op_{NC} $ have magnitude

\beq
\left|\B \nu_e^O |\Op_{NC} |\nu_\x^O \R\right| =
\frac{|(U_{e1}^* U_{\x1} + U_{e2}^* U_{\x2})
      (-U_{\x1}^* U^*_{e2} + U_{\x2}^* U^*_{e1})|}
     {|U_{e1}|^2+|U_{e2}|^2}
\simeq |U_{eh}^* U_{\x h}|.
\label{offd}
\eeq

\sni%
The approximation uses (\ref{ortho}) and the fact that experimental
constraints, that will be discussed in Section~\ref{constraints},
imply $|U_{eh}|, |U_{\x h}|, |U_{S1}|, |U_{S2}| \ll |U_{Sh}| \simeq 1$,
from which it follows that

\beq
1 = |\mbox{det}\bU| \simeq \left|
\mbox{det}\pmatrix{U_{e1}  & U_{e2}  \cr
                   U_{\x1} & U_{\x2} } \right|
|U_{Sh}|,
\eeq

\nid%
or $|-U_{\x1} U_{e2} + U_{\x2} U_{e1}| \simeq 1$.  Note that in the
oscillation-basis $\{\nu_\K^O\}$ the operator $\Op_{NC}$ also has
non-universal diagonal couplings

\beqs
\B \nu_e^O |\Op_{NC} |\nu_e^O \R &=&
1 - |U_{eh}|^2 + \frac{|U_{eh}^* U_{\x h}|^2}{1 - |U_{eh}|^2}, \cr
\B \nu_\x^O |\Op_{NC} |\nu_\x^O \R &=&
\frac{|-U_{e2} U_{\x1} + U_{e1} U_{\x2}|^2}
     {1- |U_{eh}|^2}.
\label{diag}
\eeqs

\sni%
The off-diagonal element in (\ref{offd}) and the deviations from
unity of the diagonal matrix elements in (\ref{diag}) are small since
they are quadratic in the mixing between the doublet and singlet
neutrinos.


\subsection{NP-coupling of the $W$}
The couplings of the $W$-boson are also affected by the presence of a
heavy singlet neutrino.  The effective four-Fermi couplings for
$W$-mediated neutrino scattering in the sun or a supernova are
proportional to the Charged Current (CC) operator

\beq
\Op_{CC} \equiv |\nu_e \R \B \nu_e|.
\eeq

\sni%
In our prefered basis $\{\nu_e^O, \nu_\x^O\}$ for the description of
neutrino oscillations its only non-vanishing matrix element is

\beq
\B \nu_e^O |\Op_{CC}|\nu_e^O \R =
\B \nu_e^O |\nu_e\R \B \nu_e|\nu_e^O \R = 1 - |U_{eh}|^2.
\eeq

\sni%
The other matrix elements vanish since $\nu_\x^O$ is orthogonal to
$\nu_e$ by definition. Thus also in our orthonormal basis $\Op_{CC}$
only projects onto the electron neutrino, but with a prefactor that is
slightly smaller than the standard one.


\subsection{Adding Triplets}
\label{triplets}
Additional $SU(2)_L$ singlets are not the only relevant extension of
the lepton sector. We briefly investigate the effects of neutrino
mixing with the neutral component of an $SU(2)_L$ triplet.  Fermionic
triplets arise naturally in the context of supersymmetric extensions
of the SM.  For example, the $SU(2)_L$ gauginos form such a triplet
with hypercharge $Y=0$. However, the neutral component has weak
isospin $t_{3L} = 0$ so it does {\it not} couple to the $Z$ and the
resulting formalism is the same as for singlets.

\spa%
The neutral component of a triplet with $Y = \pm 1$ does couple to the
$Z$.  A good example is the superpartner of the neutral component of
the Higgs triplet (with $Y = -1$) in supersymmetric left-right
symmetric models.  It has $t_{3L} = 1$ so we have to replace the
NC-operator defined in (\ref{NCop}) by

\beq
\Op_{NC}' \equiv \Op_{NC} + 2 |\nu_N \R \B \nu_N|.
\label{NCnew}
\eeq

\sni%
In the context of our simple example with two ordinary neutrinos and
one exotic, the off-diagonal element of $\Op_{NC}'$ in the
oscillation-basis is given by

\beq
 \B \nu_e^O |\Op_{NC}' |\nu_\x^O \R =  
\B \nu_e^O |\nu_N \R \B \nu_N|\nu_\x^O \R =
 \frac{ (U_{e1}^* U_{N1} + U_{e2}^* U_{N2}) 
      (-U_{N1}^* U_{e2}^* + U_{N2}^* U_{e1}^*)}
     {|U_{e1}|^2+|U_{e2}|^2},
\label{offnew}
\eeq
\sni%
which is quadratic in the mixing between ordinary and exotic
neutrinos, as in (\ref{offd}).


\subsection{Effective $Z$-induced NP-couplings}
In this section we compute the effective four-Fermi couplings,
$G_N^f$, for the neutrino flavor-changing reactions $\nu^O_e f \to
\nu^O_\ell f$ ($f=e, u, d$) mediated by the $Z$, in the presence of a
neutrino singlet.  We note that all three couplings are determined by
the single parameter 

\beq 
\varepsilon \equiv \B \nu_e^O |\Op_{NC} |\nu_\x^O \R, 
\eeq 

\sni%
which essentially gives the ratio between the $Z$-induced 
flavor-changing NC amplitudes and the usual flavor diagonal NC amplitudes.
For simplicity, we will ignore the possibility of CP violation, taking
$\varepsilon$ to be real throughout.  The flavor-changing four-Fermi
couplings follow directly from known results \cite{NCpot} for the
flavor-diagonal NCs, which are reviewed below.

\spa%
The potential $V_{NC}$ due to $Z$-mediated neutrino scattering off
the thermal background fermions can be deduced from the relevant
(Fierz rearranged) four-Fermi interaction

\beq
H_f = \frac{G_F}{\sqrt{2}} \bar f \gamma_\mu (g_V^f - \gamma_5 g_A^f) f
                          ~\bar \nu \gamma^\mu (1 - \gamma_5) \nu.
\eeq

\sni%
The vector and axial couplings are

\beqs
g_V^f &=& t_{3L}^f - 2q^f \sin^2 \theta_W \\
g_A^f &=& t_{3L}^f,
\eeqs

\sni%
where $t_{3L}^f$ is the weak isospin of the fermion $f$ and $q^f$ is
its electro-magnetic charge. We assume unpolarized background fermions
and therefore only the $\gamma_0$ component of the fermion density can
contribute.  The $\gamma_0 \gamma_5$ mixes ``small'' and ``large''
components of the spinor, so the axial coupling does not contribute.
For the large component $\gamma_0$ is 1, thus the charged fermion
coupling reduces to

\beq
\bar f \gamma_\mu(g_V^f - \gamma_5 g_A^f) f =
\delta_{\mu 0} N_f (t_{3L}^f - 2q^f \sin^2 \theta_W),
\eeq

\sni%
where $N_f$ is the fermion number density. Hence the term describing
NC interactions in the effective Hamiltonian for neutrino propagation
is

\beq
{\cal{H}}_{NC}^f= \frac{G_F N_f}{\sqrt{2}} 
 (2 t_{3L}^f - 4 q^f \sin^2 \theta_W)
 \times |\nu_\K \R \B \nu_\K|.
\eeq

\sni%
Note that the $\bar \nu \gamma^\mu \frac{1 - \gamma_5}{2} \nu$
coupling yields the unit matrix $|\nu_\K \R \B \nu_\K|$ in the basis
of weak neutrino eigenstates, which is just ${\cal O}_{NC}$.  The
factor $V_{NC}^f$ multiplying ${\cal O}_{NC}$ takes the values

\beqs
V_{NC}^e &=& G_F N_e (  4  \sin^2 \theta_W - 1) / \sqrt{2}, \\
V_{NC}^u &=& G_F N_u (-8/3 \sin^2 \theta_W + 1) / \sqrt{2}, \\
V_{NC}^d &=& G_F N_d ( 4/3 \sin^2 \theta_W - 1) / \sqrt{2}, \\
V_{NC}^p &=& G_F N_p ( -4  \sin^2 \theta_W + 1) / \sqrt{2}, \\
V_{NC}^n &=&-G_F N_n / \sqrt{2}.
\eeqs

\sni%
The contributions for the nucleons are obtained by summing the quark
potentials, i.e. $V_{NC}^p = 2 V_{NC}^u + V_{NC}^d$ and $V_{NC}^n =
V_{NC}^u + 2 V_{NC}^d$. Note that the electron and the proton
contributions cancel in electrically neutral media.

\spa%
In the presence of neutrino singlets, the effective flavor-changing
four-Fermi couplings $G_N^f$ are given by 

\beq 
G_N^f \equiv \frac{\B \nu_e^O |{\cal{H}}_{NC}^f |\nu_\x^O \R}{\sqrt{2} N_f}=
\frac{V_{NC}^f}{\sqrt{2} N_f} \varepsilon. 
\label{GNf} 
\eeq

\sni%
Equivalently, the flavor-changing NC potentials are 

\beq V_{FCNC}^f =
\varepsilon V_{NC}^f. 
\eeq 

\sni%
Note that the effective flavor-diagonal NC potentials are decreased
slightly by the deviations of the flavor-diagonal matrix elements $\B
\nu_e^O |\Op_{NC} |\nu_e^O \R$ and $\B \nu_\x^O |\Op_{NC} |\nu_\x^O
\R$ in eq.~(\ref{diag}) from unity.


\subsection{Modifications to Matter Oscillations}
\label{modif}%
The MSW mechanism~\cite{MSW} provides an elegant solution to the \SN
problem and it might also be important for neutrinos that are produced
in a supernova explosion. In Ref.~\cite{Bergmann} modifications of the
MSW solution due to flavor-changing neutrino interactions were
considered.  Although the new non-diagonal four-Fermi couplings were
suggested to stem from the exchange of new heavy particles, the
results obtained in~\cite{Bergmann} can also be used to discuss
matter-enhanced neutrino oscillations in the presence of $Z$-induced
FCNCs.

\spa%
We have shown that the projection of the known interaction eigenstates
onto the subspace spanned by the light mass eigenstates results in
flavor-changing NC neutrino interactions of strength $\varepsilon$
with respect to the flavor-diagonal NCs.  Thus, knowing the standard
NC-contributions it is straightforward to obtain the correct
expressions for the off-diagonal terms in the effective Hamiltonian
${\cal{H}}_N$ for neutrino propagation in matter

\beq
{\cal{H}}_N \pmatrix{\nu_e^O \cr \nu_\x^O} =
\frac{1}{4E}
\pmatrix{-\Delta \cos2\theta + A' & \Delta \sin 2\theta + B \cr
          \Delta \sin2\theta + B & \Delta \cos 2\theta - A' \cr}
\pmatrix{\nu_e^O \cr \nu_\x^O}.
\label{eq-of-motion}
\eeq

\sni%
Here $E$ is the neutrino energy, $\Delta \equiv m^2_2 - m^2_1$ is the
mass-squared difference of the two vacuum mass eigenstates,
$\theta$ is the vacuum mixing angle, and

\beq
A' \equiv A \left\{(1-|U_{eh}|^2) - \frac{N_n}{2N_e} \left[
1 - |U_{eh}|^2 + \frac{|U_{eh}^* U_{\x h}|^2}{1 - |U_{eh}|^2} -
\frac{|-U_{e2} U_{\x1} + U_{e1} U_{\x 2}|^2}
     {1- |U_{eh}|^2} \right] \right\}.
\eeq

\sni%
$A'$ reduces to the standard induced mass $A \equiv 2E \sqrt{2} G_F
N_e$ in the limit where the heavy singlet neutrino decouples
($|U_{S3}| \to 1$).  The parameter

\beq
B \equiv 4E \sqrt{2} (G_N^e N_e + G_N^u N_u + G_N^d N_d)
\eeq

\sni%
describes the FCNC contributions from neutrino scattering off
electrons and quarks in the sun or supernovae.  It is convenient for
later purposes to rewrite this as

\beq
B = 4E \sqrt{2} G_F N_e \epsilon (N_n /N_e), 
\label{modB} 
\eeq

\sni%
where

\beq 
\epsilon(x) \equiv \frac{1}{G_F} \left[ G_N^e + 2 G_N^u + G_N^d +
                                       (G_N^u + 2 G_N^d ) x \right]. 
\label{epsR}
\eeq

\sni%
We have used the fact that the quark densities can be expressed in
terms of $N_n $ and $N_e$ (which equals the proton density in neutral
matter).  To compute $B$ we set $\frac{B}{4E} = \sum_f \varepsilon
V_{NC}^f $ and obtain

\beq 
B = -2E \sqrt{2} G_F N_n \varepsilon \simeq 
    \pm 2E \sqrt{2} G_F N_n |U_{eh}^*U_{\x h}|.
\label{B}
\eeq

\sni%
Note that due to the cancelation of $V_{NC}^e$ and $V_{NC}^p$ the
parameter $B$ is proportional to $N_n$.  Interestingly, this implies
that the strongest possible effect arises from variation of ratio
$x=N_n/N_e$.


\section{Constraints}
\label{constraints}%
We now discuss experimental constraints on mixing between the ordinary
neutrinos and $SU(2)_L$ singlets.  A direct bound on $|U_{eh}^*
U_{\mu h}|$ comes from the KARMEN~\cite{KARMEN} experiment searching
for ``neutrino-oscillations'' in the $\bar\nu_e-\bar\nu_\mu$ channel.
The upper bound on the transition probability $P(\bar\nu_e \to
\bar\nu_\mu)$ yields

\beq
|U_{eh}^* U_{\mu h}|^2  \simeq
P(\bar\nu_e \to \bar\nu_\mu) < 3.1 \times 10^{-3} ~(90\% \mbox{~c.l.}).
\eeq

\sni%
This restricts $|U_{eh}^* U_{\mu h}|$ to be less than 0.06, which
allows for rather large FCNCs.

\spa%
Much tighter bounds on mixing with a singlet neutrino are obtained
from constraints on lepton universality, CKM unitarity, and the
measured $Z$ invisible decay width~\cite{limits}. Rather than fitting
the weak couplings of all known fermions we essentially update the
analysis of Ref.~\cite{Nardi}, which considers the simpler
possibility that only the neutrinos mix with exotic fermions.
($|U_{ah}|^2 $ is equivalent to $s^2_{\nu_a}$ in the notation 
of~\cite{Nardi}.) The ratios of the leptonic couplings 
$g_{e}, g_{\mu}, g_{\tau}$ to the $W$ are given by 

\beq 
\left( \frac{g_a}{g_b} \right)^2 = \frac{1-|U_{ah}|^2 }{1 - |U_{bh}|^2 },
 ~~~~~a, b = e, \mu, \tau. 
\eeq 

\sni%
The best test of $e-\mu$ universality of the $W$ couplings comes from
comparison of the rates for $\pi \to \mu \bar{\nu}_\mu $ and $\pi \to
e \bar{\nu}_e $, where the two most precise experiments \cite{pidecay}
can be combined~\cite{Marciano} to yield $g_\mu /g_e = 1.0012 \pm
0.0016$.  The best tests of $\mu -\tau$ universality come from ALEPH
measurements of leptonic \cite{ALEPHlep} and hadronic \cite{ALEPHhad}
$\tau$ decays.  Averaging the two determinations of $g_\tau /g_\mu $
(they are fully consistent with each other) yields $g_\tau /g_\mu =
0.9943 \pm 0.0065 $. From the leptonic decays one also obtains
\cite{ALEPHlep} $g_\tau /g_e = 0.9946 \pm 0.0064$.  The unitarity
constraint for the first row of the CKM matrix is  

\beq
\sum_{i =1,2,3} |V_{ui}|^2 = (1 - |U_{\mu h}|^2 )^{-1}. 
\eeq 

\sni%
The most recent experimental determination~\cite{Buras} is
$\sum_{i=1,2,3} |V_{ui}|^2 = 0.9972 \pm 0.0013 $, $2 \sigma$ away from
the SM value of unity.  Finally, the ratio of the $Z$ invisible decay
width to the SM prediction is given by~\cite{Nardi} 

\beq 
\frac{\Gamma_{Z \to inv}}{\Gamma_{Z\to inv}^{SM} } = 
  1 - \frac{|U_{eh}|^2}{6} -
  \frac{|U_{\mu h}|^2}{6} - \frac{ 2 |U_{\tau h}|^2}{3}.  
\eeq 

\sni%
Combining the most recent measurement of $\Gamma_{Z\to inv}$
\cite{EWWG}, obtained under the assumption of universal charged lepton
couplings to the $Z$, with the SM prediction (for $m_t = 175.6 \pm
5.5$~GeV) yields $\Gamma_{Z \to inv} / \Gamma_{Z\to inv}^{SM} = 0.995
\pm 0.004$.

\spa%
We construct a $\chi^2$ function with the experimental measurements
discussed above, and derive bounds on the mixing parameters using the
MINUIT package.  Allowing for singlet mixing with all known neutrinos
we obtain the 90\% c.l. bounds

\beq 
|U_{eh}|^2<0.0049,~~~|U_{\tau h}|^2<0.016,
\label{constraint1} 
\eeq

\sni%
and $|U_{\mu h}|^2 = - 0.0028 \pm 0.0013 $, which is $2 \sigma$ away
from the standard model but in the ``wrong direction" for singlet
mixing.  This is due to the deviation of the CKM
unitarity sum from unity.  However, we note that this does not rule
out $\nu_\mu$-singlet mixing since a discrepancy in this direction
could be accounted for by mixing of the $u$ and $d$ quarks with
$SU(2)_L$ singlets.  We therefore present a second set of more
conservative bounds (90\% c.l.) in which all known neutrinos are
allowed to mix with a singlet but the unitarity constraint has been
eliminated: 

\beq
|U_{eh}|^2 < 0.012,~~~|U_{\mu h}|^2 < 0.0096,~~~|U_{\tau h}|^2 < 0.016.
\label{constraint2} 
\eeq

\sni%
We have not included the possibility of mixing between the charged
leptons and exotics, nor have we taken into account correlations
between ALEPH's determinations of $g_{\tau} /g_{\mu}$ and $g_{\tau}
/g_e $ in \cite{ALEPHlep}.  Nevertheless we can conclude that our
parameter $\varepsilon$ should be less than about $1\%$, with the
largest values possible corresponding to conversion of $\nu_e$ to
$\nu_\tau$.


\section{Solar Neutrino Oscillations}
\label{sun}%
Using the results of Section~\ref{modif} it is straightforward to
obtain the survival probability \cite{Parke,KuoPanta,Bergmann} 
for a $\nu_e$ that was produced in the solar
center to be detected as an electron neutrino 

\beq
P_N(\nu_e \to \nu_e) = 
 \frac{1}{2} + \left( \frac{1}{2} - P_c \right) \cos 2\theta ~\cos 2\theta_N,
\label{Pee}
\eeq

\sni%
where the effective mixing is given by

\beq
\cos2\theta_N = \frac{(\Delta \cos2\theta - A')}
  {\sqrt{(\Delta \cos2\theta - A')^2 + (\Delta \sin2\theta + B)^2}}.
\label{cos-theta_N}
\eeq

\sni%
The crossing probability $P_c$ is well approximated by~\cite{Petcov}

\beq
P_c=\Theta(A_{prod}-A_{res}) \times 
    \frac{\exp \left[\pi \gamma_N F(\theta)/ 2 \right]-
          \exp \left[\pi \gamma_N F(\theta)/ 2 \sin^2 2\theta \right]}
         {1 - \exp \left[\pi \gamma_N F(\theta)/ 2 \sin^2 2\theta \right]}.
\label{Pc}
\eeq

\sni%
Here $\gamma_N$ denotes the adiabaticity parameter which is given
by~\cite{Bergmann}

\beq
\gamma_N =\frac{\Delta \sin^2 2\theta} 
               {2 E \cos2\theta ~(dN_e/dx)/N_e|_{res}} \times
               |1+2\epsilon(x_{res}) \cot 2\theta|^2,  
\label{gamma_N}
\eeq

\sni%
where $x_{res}$ is the ratio $N_n/N_p$ at the resonance.  Using these
results we have calculated the suppression rate for the three types of
solar neutrino experiments as a function of $\Delta$ and $\sin^2
2\theta$.  The calculations have been performed along similar lines to
those described in Ref.~\cite{Bergmann}. We present the individual
95\% c.l. contours (dashed for Homestake \cite{HS}, dotted for the
combined gallium experiments \cite{GALLEX,SAGE}, and solid for
Kamiokande \cite{KK}) together with the {\it combined} allowed regions
(shaded) for various positive (Fig.~\ref{AR_NP_pl}) and negative
(Fig.~\ref{AR_NP_mi}) values of $\varepsilon$.

\spa%
Although $Z$-induced FCNCs modify the individual contours at very
small $\sin^2 2\theta$, the combined ``small-angle solution'' is
(almost) unchanged for $\varepsilon=\pm 0.02$ and the large-angle
solution is not affected at all. In fact, even for larger
$\varepsilon$ there are no dramatic changes to the standard
small-angle MSW-solution.  In light of the experimental constraints on
$\varepsilon$ presented in Section~\ref{constraints} we conclude that
$Z$-induced FCNCs must have negligible impact on solar neutrino
oscillations. The cancelation of the NC potentials from electron and
proton scattering leads to effects proportional to the neutron density
$N_n$, which in our sun is at most half of the charged particle
densities $N_e=N_p$. This is unlike scenarios with flavor-changing
neutrino interactions induced by new heavy particle
exchange~\cite{Roulet,Guzzo,Barger91b,Fogli,KB97}, e.g.,
supersymmetric models without $R$-parity, where the different
contributions from scattering off quarks and electrons can add up
constructively.


\section{Supernova Neutrino Oscillations}
\label{supernova}%

Supernova explosions are intense neutrino sources. Due to the huge
supernova densities these neutrinos can be resonantly converted for
a large range of the parameters $\Delta$ and $\sin^2 2\theta$. The
impact of neutrino oscillations has been discussed extensively
in~\cite{Fuller,KuoPanSN,Jegerlehn,nu-nu}. Also it has been noted
that neutrino oscillations might help to solve the shock reheating
problem~\cite{SNshock}. In Refs.~\cite{Valle1,Valle2} resonant
neutrino conversions were studied in the presence of supersymmetric
$R$-parity violating interactions and $Z$-induced FCNCs, but for
{\it massless} neutrinos. In this Section we investigate how FCNCs
due to mixing of the known neutrinos with singlets alter the MSW
resonant conversion of {\it massive} neutrinos emerging from the
neutrino-sphere of a supernova. The object we investigate is the
survival probability $P(\nu_e \to \nu_e)$.  It has been shown in
Ref.~\cite{KuoPanSN} that $P(\nu_e \to \nu_e)$ and $P(\bar\nu_e \to
\bar\nu_e)$ are the only quantities that must be specified to
determine how neutrino oscillations mix the fluxes. We assume that
the neutrino parameters are such that matter-enhanced neutrino
oscillations can occur for the neutrinos but not the
anti-neutrinos. Moreover, we neglect neutrino-neutrino
scattering~\cite{nu-nu} which is justified for neutrino propagation
outside the neutrino-sphere. For the point we wish to make it is
sufficient to discuss the impact of \FCNC on the survival
probability itself, so we do not convolute $P(\nu_e \to \nu_e)$
with the predicted neutrino fluxes and cross sections in order to
compute the experimental rates. (Note that from the neutrino data
of SN1987a \cite{SNobs} one cannot obtain a reasonable energy
spectrum.  Moreover, it might well be that NP processes, like
$Z$-induced FCNCs, have an effect on the flux of produced neutrinos, 
which has not been considered in current supernova 
simulations~\cite{SNtheory}.)

\spa%
As shown in Ref.~\cite{KuoPanSN}, the following modifications of the
solar MSW formalism are required in the case of supernova neutrinos:

\sni%
(a) The electron-density at the neutrino-sphere $(N_e)_{prod} \sim
10^{35}$ cm$^{-3}$ is larger than the solar core density $(N_e)_{core}
\sim 10^{25}$ cm$^{-3}$ by about ten orders of magnitude. Since the
supernova neutrinos are not much more energetic than solar neutrinos,
an adiabatic threshold energy

\beq
E_A \equiv \frac{\Delta \cos 2\theta}{2\sqrt{2} G_F (N_e)_{prod}}
\label{E_A}
\eeq

\sni%
of the order of a few MeV will scale the adiabatic band by a factor 
$\sim 10^{10}$ with respect to the solar case, to $\Delta_{max} 
\sim 10^6$ eV$^2$. Above this value neutrinos cannot be resonantly 
converted, since they are produced below the resonance. Note that 
$E_A$ is not changed by flavor-changing neutrino 
interactions~\cite{Bergmann}.

\sni%
(b) Unlike the solar density profile, which decays (roughly)
exponentially, the supernova density $\rho(r)$ is predicted to
decrease like $1/r^3$ ($r$ being the distance to the core outside the
neutrino-sphere), but for the sake of generality we will just assume a
power law (with $\alpha > 0$):

\beq
\rho(r) = \rho(R) \left(\frac{r}{R}\right)^{-\alpha}.
\label{power}
\eeq

\sni%
For a density described by (\ref{power}) the scaling factor $N_e'/N_e$
that appears in the adiabaticity factor $\gamma_N$ [defined in
Eq.~(\ref{gamma_N})] is not constant (like for an exponentially
decaying density profile), but:

\beq
\frac{N_e'}{N_e} = \frac{-\alpha}{r}
                 = \left(\frac{-\alpha}{R}\right)
                   \left[\frac{N_e(r)}{N_e(R)}\right]^{1/\alpha},
\label{scale}
\eeq

\sni%
where we assume that the electron number density $N_e=\rho Y_e/m_N$ ($m_N$
is the nucleon mass) is proportional to the mass density $\rho$ (in
fact $Y_e \simeq 0.4$ is constant to a good approximation outside the
neutrino-sphere). To obtain $\gamma_N$ we have to take $N_e$ in
(\ref{scale}) at the resonance

\beq
N_e^{res}=\frac{\Delta \cos 2\theta}{2\sqrt{2} G_F E}.
\label{Nres}
\eeq

\sni%
resulting in

\beqs
\gamma_N &=& \frac{\sin^2 2\theta |1+2\epsilon(x_{res}) \cot 2\theta|^2}
                  {2E \Delta \cos 2\theta}
\frac{R}{\alpha}
\left(\frac{2\sqrt{2} G_F E N_e(R)}{\Delta \cos 2\theta}\right)^{1/\alpha} 
\nonumber\\
&=& E^\frac{1-\alpha}{\alpha} ~\Delta^\frac{\alpha-1}{\alpha}
    ~(\cos 2\theta)^{-\frac{\alpha+1}{\alpha}}
    ~|\sin 2\theta + 2\epsilon(x_{res}) \cos 2\theta|^2 \nonumber \\
&~& \times \frac{R}{2\alpha} \left[2\sqrt{2} G_F N_e(R)\right]^{1/\alpha}.
\eeqs

\sni%
Thus, the effective non-adiabatic threshold energy is

\beqs
E_{NA} &=& \Delta ~(\cos 2\theta)^{-\frac{\alpha+1}{\alpha-1}}
~|\sin 2\theta + 2\epsilon(x_{res}) \cos 2\theta|^\frac{2\alpha}{\alpha-1}
\times \left[ 2\sqrt{2} G_F N_e(R) \right]^\frac{1}{\alpha-1}
\times \left(\frac{\pi R}{4\alpha}\right)^\frac{\alpha}{\alpha-1} 
\nonumber \\ 
&=& 3.7 \times 10^9 ~\mbox{MeV} 
~\left(\frac{\Delta}{\mbox{eV}^2}\right) \cos^{-2} 2\theta
~|\sin 2\theta + 2\epsilon(x_{res}) \cos 2\theta|^3,
\label{E_NA}
\eeqs

\sni%
where for the last line we have taken typical supernova values, i.e.,
$R=10^7$ cm for the radius of the neutrinosphere, $\rho(R)=10^{12}$
g/cm$^3$ for the density at $R$ and $\alpha=3$. From (\ref{E_NA}) it
is clear that the non-adiabatic band starts off at similar values of
$\Delta_{min} \sim 10^{-9}$ eV$^2$ as in the sun, but has a different
slope of~$-3/2$ (for $\epsilon(x_{res})=0$) in the (logarithmic)
$\Delta - \sin^2 2\theta$ plane.

\sni%
(c) Since the central supernova density is so huge, the ``higher''
$e-\tau$ resonance almost always precedes the ``lower'' $e-\mu$
resonance. Thus the authors of Ref.~\cite{KuoPanSN} have pointed out
that a proper treatment of supernova neutrino oscillations should be
done within a 3-flavor formalism. Moreover they have noted
in~\cite{KuoPan3nu} that for small mixing angles the survival
probability factorizes into

\beq
P(\nu_e \to \nu_e) = P_l(\nu_e \to \nu_e) \times P_h(\nu_e \to \nu_e),
\label{Plh}
\eeq

\sni%
where $P_{l,h}$ are the standard two-level survival probabilities for
the lower and higher resonances, respectively. Since flavor-changing
neutrino interactions become important when the mixing is small (i.e.
$\tan 2\theta \lsim |\epsilon|$), eq.~(\ref{Plh}) is sufficient for
our analysis. We will assume that $\Delta_l$ and $\sin^2 2\theta_l$ of
the lower resonance are fixed by the standard MSW-solution to the \SN
problem (rather than fixing the ratios of $\Delta_l/\Delta_h$ and
$\sin^2 2\theta_l/\sin^2 2\theta_h$ at some arbitrary value) in order
to obtain a two-dimensional plot.  Then, at fixed energy $E_\nu$, the
survival probability $P(\nu_e \to \nu_e)$ is just a constant $P_l$
multiplying $P_h(E_\nu,\sin 2\theta_h, \Delta_h, \varepsilon)$ which
we show in Fig.~\ref{SN0}.  The plot exhibits the features of the
supernova ``MSW-triangle'' that we discussed in (a) and (b): The
adiabatic band appears at very large $\Delta_{max} \sim 10^6$ eV$^2$
and the non-adiabatic band starts off at $\Delta_{min} \sim
10^{-9}$ eV$^2$ for $\sin^2 2\theta=1$ extending to very small mixing
$\sin^2 2\theta \sim 10^{-10}$ where it connects to the adiabatic
band.

\spa%
The important consequence of this is that the supernova triangle is
sensitive to even tiny FCNCs.  One can see this clearly in
Fig.~\ref{NP-monop} and Fig.~\ref{NP-monom} where we show the survival
probability for various positive and negative $\varepsilon$.  The
effect can be easily understood in terms of the adiabaticity parameter
$\gamma_N$. Without \FCNC $\gamma \propto \Delta^{(2/3)} \sin^2
2\theta$ for small vacuum mixing. Thus for smaller $\Delta$ (and fixed
$\sin^2 2\theta$) the propagation is less adiabatic and there will be
a minimal value $\Delta_{min}$ for each value of $\sin^2 2\theta$
where most of the neutrinos ``cross-over'', resulting in a large
survival probability. The non-adiabatic band is the contour defined by
$\Delta_{min}(\sin^2 2\theta)$ which separates the regions where the
adiabatic conversion is efficient (above) and where it is not (below).
As we have mentioned, this band is a straight line (with slope~$-3/2$)
if there are no \FCNC. However, if $\epsilon \ne 0$ then the
adiabaticity parameter behaves like $\gamma_N \propto \Delta^{(2/3)}
|\sin 2\theta + 2\epsilon|^2$ for small vacuum mixing.  Thus for $\sin
2\theta \ll |\epsilon|$, $\Delta_{min}$ is determined by $\epsilon$,
the strength of the \FCNC, and {\it not} by the vacuum mixing $\sin
2\theta$ as can be seen in Fig.~\ref{NP-monop} and
Fig.~\ref{NP-monom}.  Simply, in this regime the off-diagonal elements
of the effective Hamiltonian in (\ref{eq-of-motion}) are dominated by
the FCNC term rather than the mixing term.  Note that for positive
$\varepsilon$ (corresponding to negative $\epsilon$) the two competing
contributions from \FCNC and mixing can cancel each other resulting in
a vanishing $\gamma_N$ which implies a large survival probability
around $\sin 2\theta_{div} = -2\epsilon$ (for a more detailed
discussion see Ref.~\cite{Bergmann}).  The important result is that
neutrino FCNC effects can be very significant for supernova neutrinos.


\section{Conclusions}
\label{conclusions}
We have argued that neutrino singlets are almost unavoidable in any
framework that attempts to solve neutrino anomalies by neutrino
oscillations. The additional heavy neutrinos give rise to $Z$-induced
FCNC interactions in the effective matter propagation matrix, and the
question of whether or not they can be neglected is rather a
quantitative than qualitative one. We have worked out the resulting
modifications to the MSW mechanism in order to study FCNC effects on
matter-enhanced neutrino oscillations in the sun and in supernovae.
We have found that while phenomenological constraints rule out
significant changes in the \SN MSW-solutions, the impact of \FCNC on
the survival probability of supernova neutrinos can be large, even for
very small singlet components in the standard neutrino mass
eigenstates.

\spa%
We conclude by asking whether values of $\varepsilon$ which are large
enough to be of relevance to supernova neutrinos naturally occur in
scenarios employing a neutrino see-saw mechanism.  For example, this
is easily seen not to be the case for the original see-saw matrix in
eq.~(\ref{see-saw-matrix}).  The singlet-doublet mixing is of order
$m_D/m_R$ so that $\varepsilon \sim m_D^2 / m_R^2$.  This should be
$\gsim 10^{-5}$ for light-heavy mixing effects to be relevant for
supernova neutrinos. But in this case the light neutrino masses would
only be suppressed by a factor $m_D/m_R \gsim 10^{-3}$ with respect to
the Dirac masses. If the latter are reasonably large, e.g., of the
order of the charged lepton masses, this would be inconsistent with
the ultra-light neutrinos usually required to explain the \SN problem
(although strictly speaking only $\Delta$ has to be tiny).

\spa%
However, as already noted in the introduction, there are variants of
the above scenario which can lead to much larger singlet admixtures.
In a popular alternative, the Majorana ``double see-saw''
\cite{Gonzalez}, two singlets are added (per generation) leading to a
$9\times 9$ mass matrix of the form 

\beq 
{\bM'} = \pmatrix{0       & \bM_D   & 0     \cr 
                  \bM_D^T & 0       & \bM_R \cr 
                  0       & \bM_R^T & \bM_S \cr}.
\label{see-saw-matrix2}
\eeq

\sni%
For $m_D, m_S \ll m_R$ ($m_X \equiv (\mbox{det} \bM_X)^{1/3}$)
singlet-doublet mixing is still of order $m_D/m_R$ but the light
neutrino masses are of order $m_S (m_D /m_R)^2$.  Hence in this
framework one can obtain ultra-light neutrinos while having a ratio
$m_D /m_R \gsim 10^{-3}$ that induces \FCNC that are significant for
the MSW-effect in supernovae.  Since the Dirac masses of the neutrinos
have to be smaller than the weak scale, $m_{weak}$, it follows that
as long as the mixing is $\sim m_D /m_R$ the $Z$-induced \FCNC are
potentially relevant to our discussion if $m_R \lsim 100$ TeV.

\spa%
Finally, vectorlike pairs of $SU(2)_L$ singlet quarks and leptons with
large masses $m_V$ are often introduced in order to suppress known
quark and charged lepton masses relative to the weak scale via a
generalized ``Dirac see-saw'' \cite{D-seesaw}, leading to
left-handed singlet components in the ordinary charged fermion mass
eigenstates of order $m_{weak} /m_V $.  If vectorlike pairs of
neutral singlets are also present in such a scenario the known
neutrinos could be expected to mix with the left-handed singlets at
same order as the charged fermions, in addition to mixing with
right-handed singlets responsible for an ultra-light Majorana mass
see-saw. (In the absence of a Majorana see-saw the vectorlike singlets
would typically lead to neutrino Dirac masses which are of same order
as the charged lepton masses.) As in the above example, singlet mixing
at the level of interest for supernova oscillations, i.e., $m_{weak}
/m_V \gsim 10^{-3}$, would correspond to a New Physics mass scale $m_V
\lsim 100$ TeV.  The right-handed neutrino Majorana mass scale $m_R$
could be arbitrarily large \cite{GrossNir}, thus allowing for ultra-light 
neutrino masses which are consistent with the MSW solution for solar 
neutrinos.

\sni%
{\it Note added:} When this work was near completion we learned of
another paper \cite{KuoMan} that analyzed the effects of \FCNC on
supernova oscillations in the context of supersymmetric models with
broken $R$-parity, arriving at effects of similar magnitude to those
presented in our analysis.

\spa%
\spa%
\spa%
\begin{center}
\bf Acknowledgments
\end{center}

\sni%
We thank Y. Nir and Y. Grossman for useful discussions and valuable
comments on the manuscript.  We are also grateful to M. Sokoloff for
help with the MINUIT package.  A.K. would like to thank the Weizmann
Institute Physics Department for a very enjoyable visit while this
work was in progress.  A.K. was supported by the United States
Department of Energy under Grant No. DE-FG02-94ER40869.


{}


\putMSW{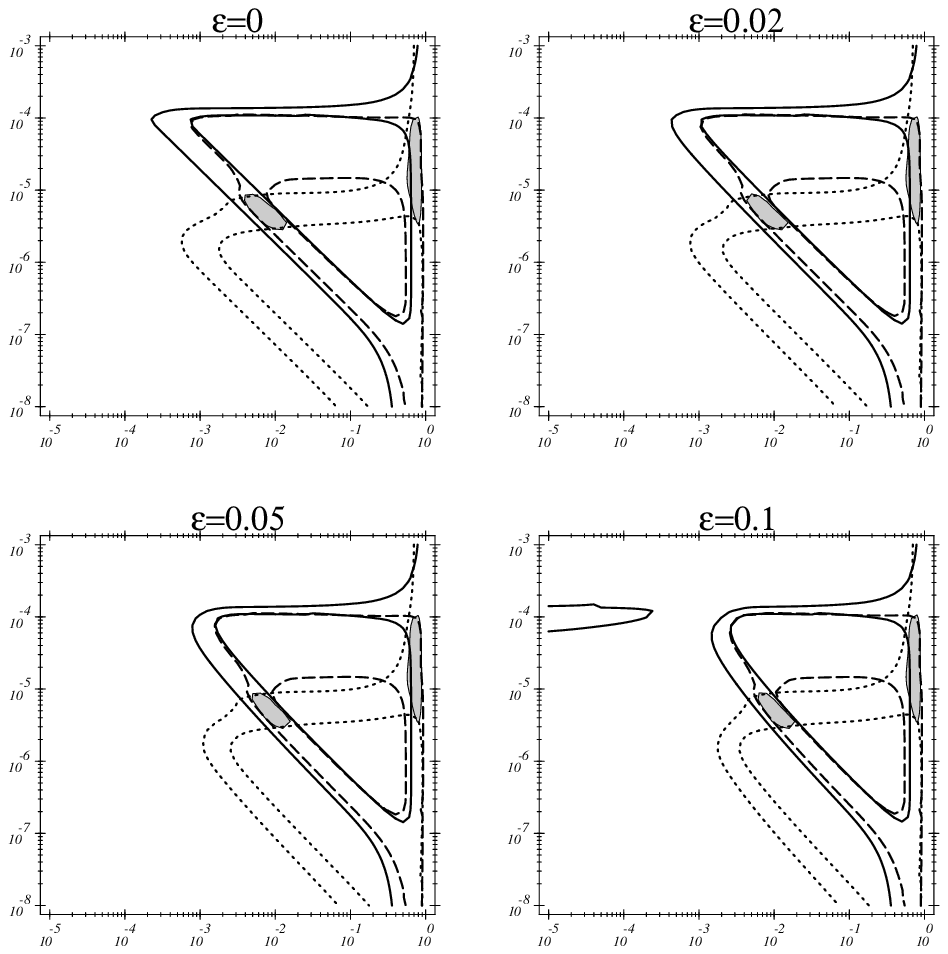}{The combined allowed regions for the solar neutrino
  experiments with $\varepsilon \ge 0$ as indicated above each plot. The
  dotted contours correspond to the combined gallium experiments, the
  dashed contours to the Homestake experiment and the solid contours
  to the Kamiokande experiment. The shaded areas indicate the 95\%
  c.l.  combined allowed regions in the $\sin^2 2\theta - \Delta$
  plane.}{\widthAR}{\heightAR}{\spaAR}

\putMSW{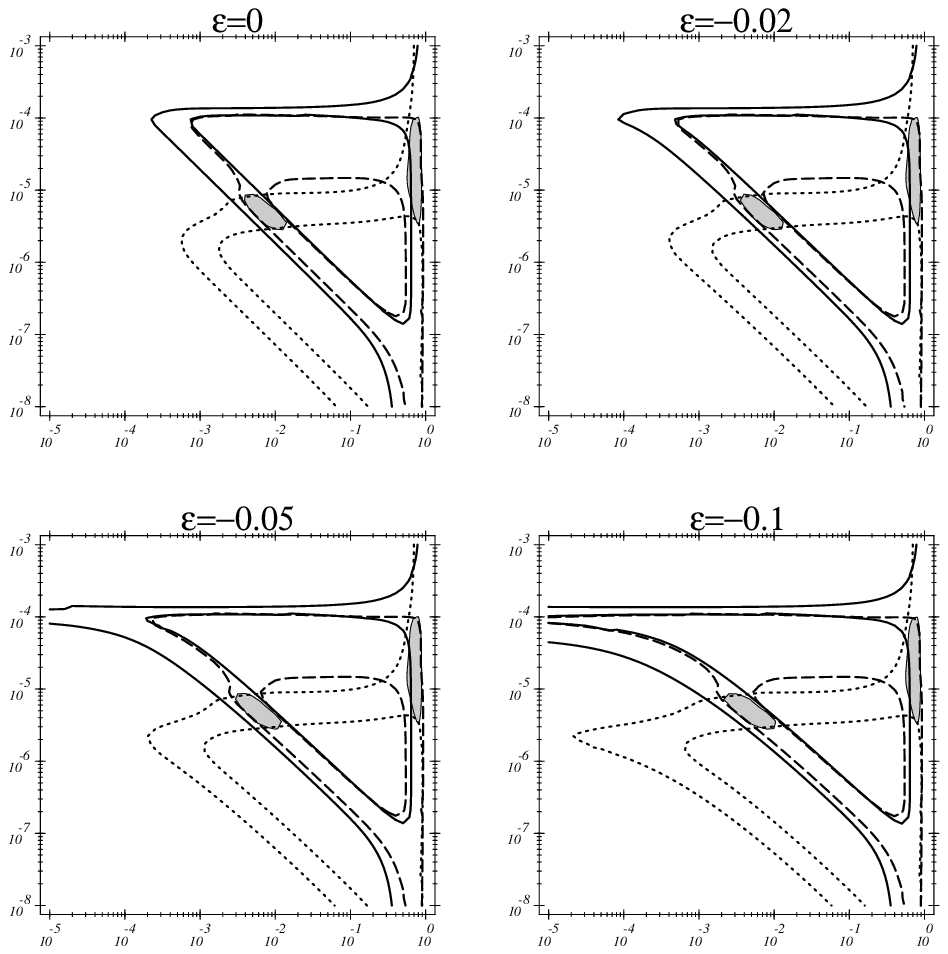}{The combined allowed regions for the solar neutrino
  experiments with $\varepsilon \le 0$ as indicated above each plot. The
  dotted contours correspond to the combined gallium experiments, the
  dashed contours to the Homestake experiment and the solid contours
  to the Kamiokande experiment. The shaded areas indicate the 95\%
  c.l.  combined allowed regions in the $\sin^2 2\theta - \Delta$
  plane.}{\widthAR}{\heightAR}{\spaAR}

\putMSW{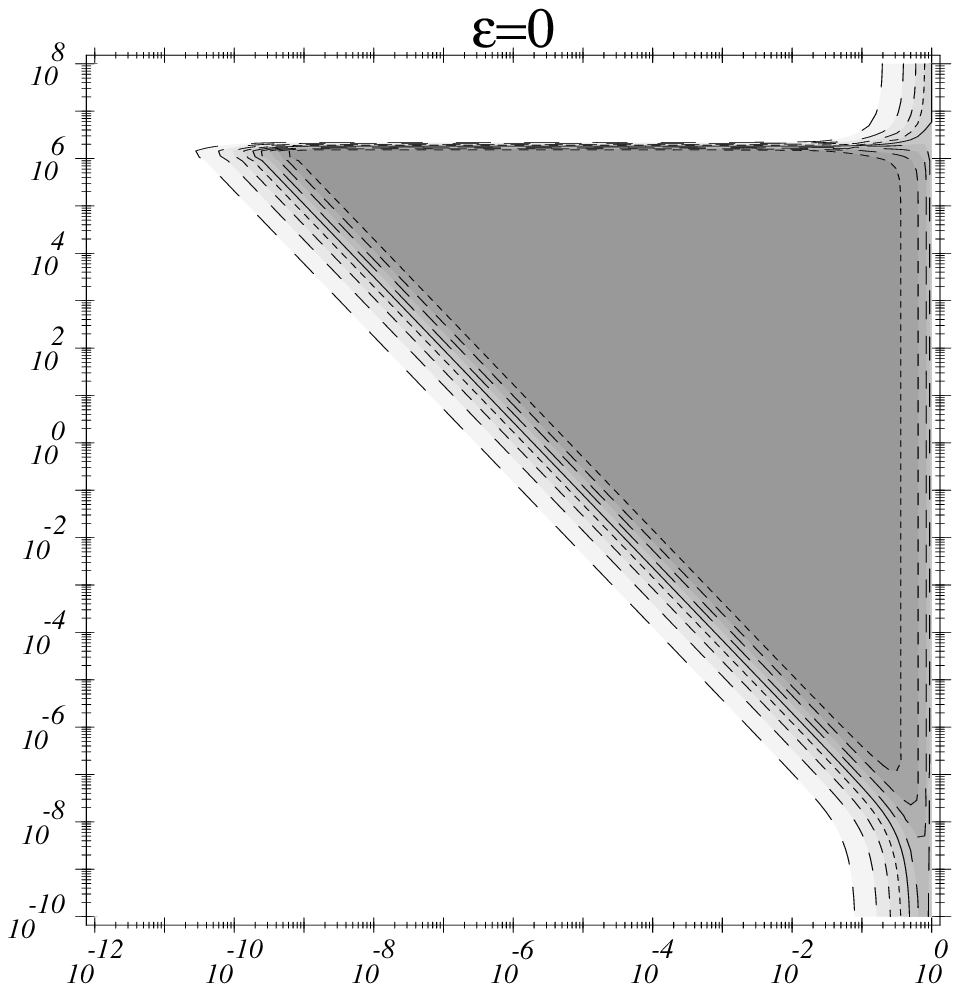}{The MSW-contours for supernova neutrinos with
  $\varepsilon = 0$ at one discrete energy ($E_\nu = 10$~MeV). The
  shading indicates the value of the survival probability $P_h (\nu_e
  \to \nu_e)$ in the $\sin^2 2\theta - \Delta$ plane: White
  corresponds to $0.9 \le P_h \le 1.0$ and the darkest area
  corresponds to $0.0 \le P_h \le 0.1$.}{\heightAR}{\widthAR}{\spaAR}

\putMSW{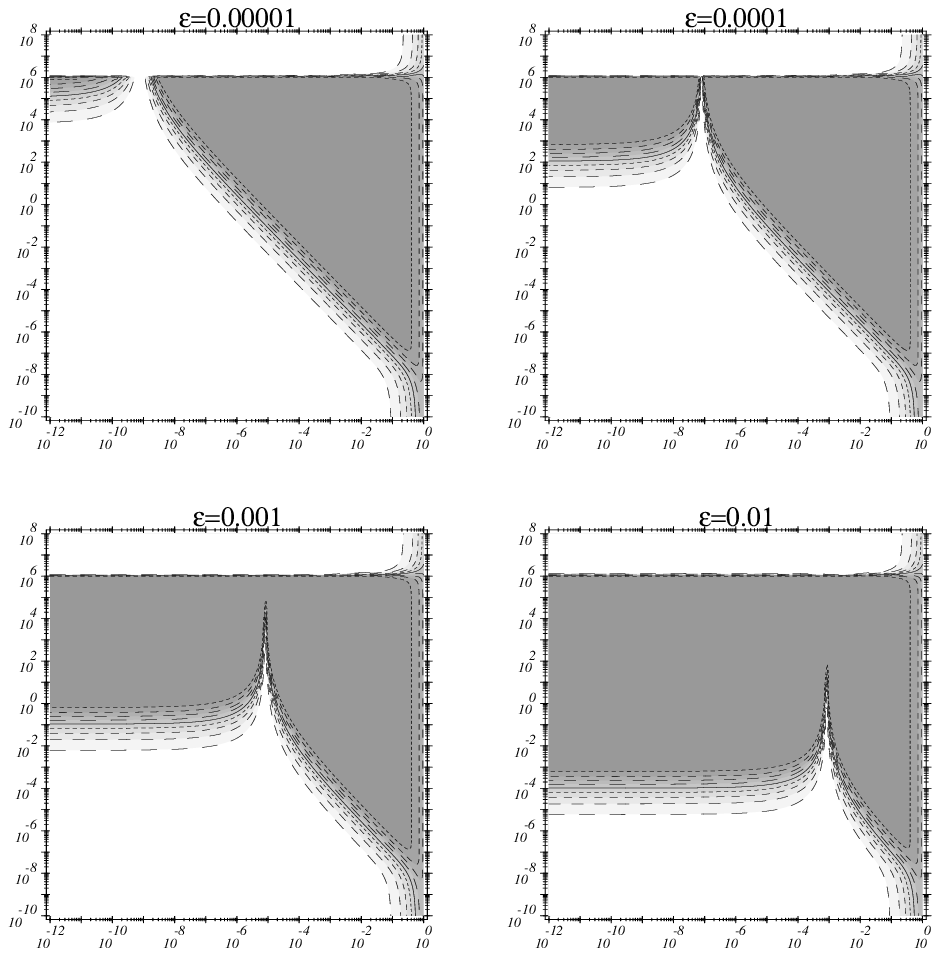}{The MSW-contours for supernova neutrinos with
  $\varepsilon > 0$ (indicated above each plot) at one discrete energy
  ($E_\nu = 10$~MeV). The shading indicates the value of the survival
  probability $P_h(\nu_e \to \nu_e)$ in the $\sin^2 2\theta - \Delta$
  plane: White corresponds to $0.9 \le P_h \le 1.0$ and the darkest
  area corresponds to $0.0 \le P_h \le
  0.1$.}{\heightAR}{\widthAR}{\spaAR}

\putMSW{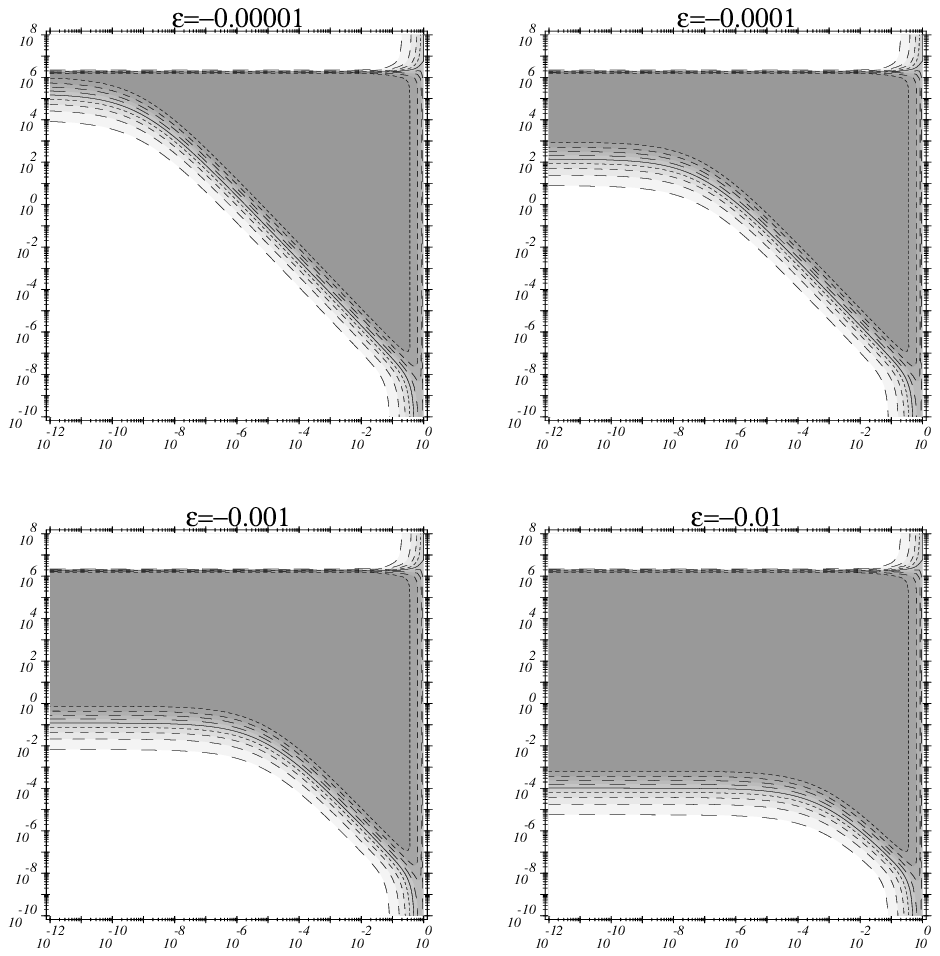}{The MSW-contours for supernova neutrinos with
  $\varepsilon < 0$ (indicated above each plot) at one discrete energy
  ($E_\nu = 10$~MeV). The shading indicates the value of the survival
  probability $P_h(\nu_e \to \nu_e)$ in the $\sin^2 2\theta - \Delta$
  plane: White corresponds to $0.9 \le P_h \le 1.0$ and the darkest
  area corresponds to $0.0 \le P_h \le
  0.1$.}{\heightAR}{\widthAR}{\spaAR}


\newpage
\begin{thebibliography}{99}
\bibitem{KuoPanta}  For a review on neutrino oscillations see:\\
                    T. K. Kuo and J. Pantaleone,
                    Rev. Mod. Phys. 61 (1989) 937.

\bibitem{HS}        R. Davis {\it et al.}, \PRL 20, (1968) 1205; \\
                    R. Davis, Prog. Part. Nucl. Phys. 32 (1994) 13; \\
                    B. T. Cleveland et al.,
                     \NPB (Proc. Suppl.) 38 (1995) 47; \\
                    K. Lande, to be published in {\it Neutrino 96},
                     Proceedings of the 17th International Conference
                     on Neutrino Physics and Astrophysics,
                     Helsinki, Finland, 13--19 June 1996,
                     edited by K. Huitu, K. Enqvist and J. Maalampi
                     (World Scientific, Singapore).

\bibitem{GALLEX}    P. Anselmann {\it et al.}, \\
                     \PLB 285 (1992) 390, 327 (1994) 377,
                          342 (1995) 440, 357 (1995) 237; \\
                    T. Kirsten {\it et al.} (GALLEX Collaboration),
                     to be published in {\it Neutrino 96},
                     Proceedings of the 17th International Conference
                     on Neutrino Physics and Astrophysics,
                     Helsinki, Finland, 13--19 June 1996,
                     edited by K. Huitu, K. Enqvist and J. Maalampi
                     (World Scientific, Singapore).

\bibitem{SAGE}      A. I. Abasov {\it et al.}, \PRL 67 (1991) 3332; \\
                    J. N. Abdurashitov {\it et al.},
                     \PLB 328 (1994) 234; \\
                    V. Gavrin {\it et al.} (SAGE Collaboration),
                     to be published in {\it Neutrino 96},
                     Proceedings of the 17th International Conference
                     on Neutrino Physics and Astrophysics,
                     Helsinki, Finland, 13--19 June 1996,
                     edited by K. Huitu, K. Enqvist and J. Maalampi
                     (World Scientific, Singapore).

\bibitem{KK}        K. S. Hirata {\it et al.}, \PRL 62 (1989) 16; \\
                    Y. Fukuda {\it et al.} (Kamiokande Collaboration),
                     \PRL 77 (1996) 1683.

\bibitem{SSM}       J. N. Bahcall and M. H. Pinsonneault,
                     Rev. Mod. Phys. 67 (1995) 781;\\
                    see also:
                    J. N. Bahcall, {\it Neutrino Astrophysics}
                    (Cambridge University Press, Cambridge, England, 1989).

\bibitem{Hata}      N. Hata and P. Langacker, \PRD 56 (1997) 6107
                     (hep-ph/9705339).

\bibitem{helioseis} J. N. Bahcall, M. H. Pinsonneault, S. Basu,
                     J. Christensen-Dalsgaard, \\
                    \PRL 78 (1997) 171 (astro-ph/9610250).

\bibitem{MSW}       L. Wolfenstein, \PRD 17 (1978) 2369;\\
                    S. P. Mikheyev and A. Yu. Smirnov,
                     Yad Fiz. 42 (1985) 1441.

\bibitem{PDG}       R. M. Barnett {\it et al.}, Particle Data Group,
                     \PRD 54 (1996).

\bibitem{seesaw}    T. Yanagida, {\it Proceedings of the Workshop on
                    Unified Theory and Baryon Number in the Universe},
                    (KEK, Japan, 1979); \\
                    M. Gell-Mann, R. Slansky and P. Ramond,
                    {\it Supergravity}, (North Holland 1979) 346.

\bibitem{ANproblem} M. C. Gonzalez-Garcia, H. Nunokawa, O. Peres,
                    T. Stanev, J. W. F. Valle,\\ hep-ph/9712238; \\
                    see also references therein and:
                    B. Barish, \NPB (Proc. Suppl.) 38 (1995) 343.

\bibitem{LSNDdar}   C. Athanassopoulos {\it et al.}, \PRL 77 (1996) 3082.

\bibitem{LSNDdif}   C. Athanassopoulos {\it et al.},  nucl-ex/9706006.

\bibitem{Z-FCNC}    P. Langacker and D. London, \PRD 38 (1988) 907; \\
                    H. Nunokawa, Y.-Z. Qian, A. Rossi and J. W. F. Valle, \\
                     \PRD 54 (1996) 4356 (hep-ph/9605301); \\
                    see also references therein.

\bibitem{Schechter} J. Schechter and J. W. F. Valle, \PRD 22 (1980) 2227.

\bibitem{NCpot}     D. N\"{o}tzold and G. Raffelt, \NPB 307 (1988) 924;\\
                     see also Ref.~\cite{KuoPanta}.

\bibitem{Bergmann}  S. Bergmann, hep-ph/9707398,
                     \NPB, in press.

\bibitem{KARMEN}    KARMEN-Collaboration,
                     \NPB (Proc. Suppl.) 38 (1995) 235; \\
                    see also hep-ex/9706023.

\bibitem{limits}    E. Nardi, E. Roulet and D. Tommasini, 
                     \PLB 344 (1995) 225;\\
                    C. P. Burgess {\it et al.}, \PRD 49 (1994) 6115;\\
                    P. Langacker and D. London, \PRD 38 (1988) 886.

\bibitem{Nardi}     E. Nardi, E. Roulet and D. Tommasini, \PLB 327 (1994) 319.

\bibitem{pidecay}   D. I. Britton {\it et. al.}, \PRL 68 (1992) 3000; \\
                    C. Czaapek {\it et. al.}, \PRL 70 (1993) 17.

\bibitem{Marciano}  W. Marciano, Proc. of the 3rd Workshop 
                    on Tau Lepton Physics, Montreux 1994; \\ 
                    L. Rolandi ed., {\NPB (Proc. Suppl.)} (1995).

\bibitem{ALEPHlep}  ALEPH Collaboration, Z. Phys. C 70 (1996) 561.

\bibitem{ALEPHhad}  ALEPH Collaboration, Z. Phys. C 70 (1996) 579.

\bibitem{Buras}     A.J. Buras, Proceedings of the 7th International 
                    Symposium on Heavy Flavor Physics, 
                    Santa Barbara, July 1997.

\bibitem{EWWG}      LEP Electroweak Working Group and SLD Heavy Flavour Group,
                    LEPEWWG/97-02.

\bibitem{Parke}     S. J. Parke, \PRL 57 (1986) 1275.

\bibitem{Petcov}    S. T. Petcov, \PLB 200 (1988) 373; \\
                     see also: \NPB (Proc. Suppl.) 13 (1990) 527.

\bibitem{Roulet}    E. Roulet, \PRD 44 (1991) 935.
                    
\bibitem{Guzzo}     M. M. Guzzo, M. Masiero and S. T. Petcov,
                     \PLB 260 (1991) 154; \\
                    M. M. Guzzo and S. T. Petcov,
                     \PLB 271 (1991) 172.

\bibitem{Barger91b} V. Barger, R. J. N. Phillips and K. Whisnant,
                     \PRD 44 (1991) 1629.

\bibitem{Fogli}     G. L. Fogli and E. Lisi,
                     Astroparticle Physics 2 (1994) 91.

\bibitem{KB97}      P. I. Krastev and J. N. Bahcall,
                     hep-ph/9703267. 
                
\bibitem{Fuller}    G. M. Fuller, R. W. Mayle and J. R. Wilson,
                     \APJ 322 (1987) 795.

\bibitem{KuoPanSN}  T. K. Kuo and J. Pantaleone, \PRD 37 (1988) 298.

\bibitem{Jegerlehn} B. Jegerlehner, F. Neubig and G. Raffelt,
                    \PRD 54 (1996) (astro-ph/9601111).

\bibitem{nu-nu}     J. Pantaleone, \PLB 342 (1995) 250 (astro-ph/9405008); \\
                    Y.-Z. Qian and G. M. Fuller,
                     \PRD 51 (1995) 147 (astro-ph/9406073). 

\bibitem{SNshock}   G. M. Fuller, R. W. Mayle B. S. Meyer and J. R. Wilson, \\ 
                    \APJ 389 (1992) 517.

\bibitem{Valle1}    H. Nunokawa, A. Rossi and J. W. F. Valle, \\
                    \NPB 482 (1996) 481 (hep-ph/9606445).

\bibitem{Valle2}    H. Nunokawa, Y.-Z. Qian, A. Rossi and J. W. F. Valle, \\
                    \PRD 54 (1996) 4356 (hep-ph/9605301).

\bibitem{SNobs}     K. Hirata {\it et al.},
                     \PRL 58 (1987) 1490; \\
                    R. M. Bionta {\it et al.},
                     \PRL 58 (1987) 1494.

\bibitem{SNtheory}  S. E. Woosley, J. R. Wilson and R. Mayle,
                     \APJ 302 (1986) 19; \\
                    R. Mayle, J. R. Wilson and D. N. Schramm,
                     \APJ 318 (1987) 288; \\
                    A. Burrows, D. Klein and R. Gandhi,
                     \PRD 45 (1992) 3361.

\bibitem{KuoPan3nu} T. K. Kuo and J. Pantaleone,
                     \PRD 35 (1987) 3432.

\bibitem{Gonzalez}  R. N. Mohapatra, \PRL 56 (1986) 561; \\
                    M. C. Gonzalez-Garcia and J. W. F Valle,
                     \PLB 216 (1989) 360; \\
                    I. Antoniadis and G.K. Leontaris, \PLB 216 (1989) 333.

\bibitem{D-seesaw}  See, for example, 
                    C. D. Froggatt and H. B. Nielsen, \NPB 147 (1979) 277;\\
                    A. Davidson and K. C. Wali, \PRL 59 (1987) 393;\\
                    M. Leurer, Y. Nir and N. Seiberg, 
                                         \NPB 398 (1993) 319, 420 (1994) 468.

\bibitem{GrossNir}  Y. Grossman and Y. Nir, \NPB 448 (1995) 30. 

\bibitem{KuoMan}    S. W. Mansour and T. K. Kuo, hep-ph/9711424.

\end{thebibliography}
\end{document}